\documentstyle[11pt,paspconf,epsf]{article}

\begin{document}

\title{Cosmology With High-Redshift Water Masers}
\author{J. R. Herrnstein}
\affil{National Radio Astronomy Observatory, Socorro, NM 87801}

\begin{abstract}
Multi-epoch VLBA observations of the maser in NGC 4258 have yielded a 
4\% geometrical distance to the galaxy.  The potential scientific 
payoffs of finding similar objects at large distances, in the Hubble 
flow, are considerable.  In this contribution, I discuss the 
plausibility of detecting high-redshift water masers, and describe
a search strategy that we have implemented to realize this objective. 
\end{abstract}


\section{Introduction -- The Maser in NGC 4258}

VLBA observations of the water masers in NGC 4258 reveal a nearly edge-on, slightly 
warped, extremely thin disk in nearly perfect Keplerian rotation around a central 
binding mass of $3.5\times10^{7}$\,M$_{\odot}$ (Watson \& Wallin 1994; Greenhill {\it et al.}
1995; Miyoshi {\it et al.} 1995; Moran {\it et al.} 1995; Herrnstein, Greenhill, \& Moran 1996).
The VLBA observations of the NGC 4258 maser provide insight into the structure and 
kinematics of the accretion disk, and NGC 4258 is an exceptional laboratory for the 
study of AGN accretion phenomenon and the connection between accretion disks and 
jet emission (Herrnstein {\it et al.} 1997; Herrnstein {\it et al.} 1998a). They can also be used 
to derive a precise geometric distance to NGC 4258.  

The upper panel of Figure~1 is a schematic representation of the best-fitting NGC~4258 
disk model, as derived from the positions and line-of-sight (LOS) velocities of the 
masers.  As the disk rotates, the 'systemic' masers along the near edge of the disk 
drift in position and LOS velocity by about 30\,$\mu$as\,yr$^{-1}$ and 9\,km\,s$^{-1}$, respectively, 
with respect to the apparently stationary 'high-velocity' masers in the plane of the 
sky (Herrnstein 1997a\&b). The expressions at the bottom of Figure~1 illustrate that
{\it both} the LOS accelerations ($\dot{v}_{LOS}$) {\it and} the proper motions ($\dot{\theta}_{x}$) 
can be used to derive a purely geometric distance ($D$) to NGC 4258.  Because the actual
space velocities ($v_{rot}$) and angular radii ($\theta_{R}$) of the systemic masers cannot 
be measured directly, these acceleration and proper motion distances are model dependent.  
Fortunately, however, most of the model dependence resides in $\theta_{R}$, and the two 
distance estimates together provide a geometric distance estimate that is largely 
model independent. With this in mind, we have observed NGC 4258 with the VLBA at 
3--4 month intervals for the last three years.  The LOS accelerations and proper motions 
provided by the first five epochs of these data yield a geometric distance of 
$7.3\pm0.3$\,Mpc (Herrnstein 1997a; Herrnstein {\it et al.} 1998b).  

\begin{figure}[p]     
\plotone{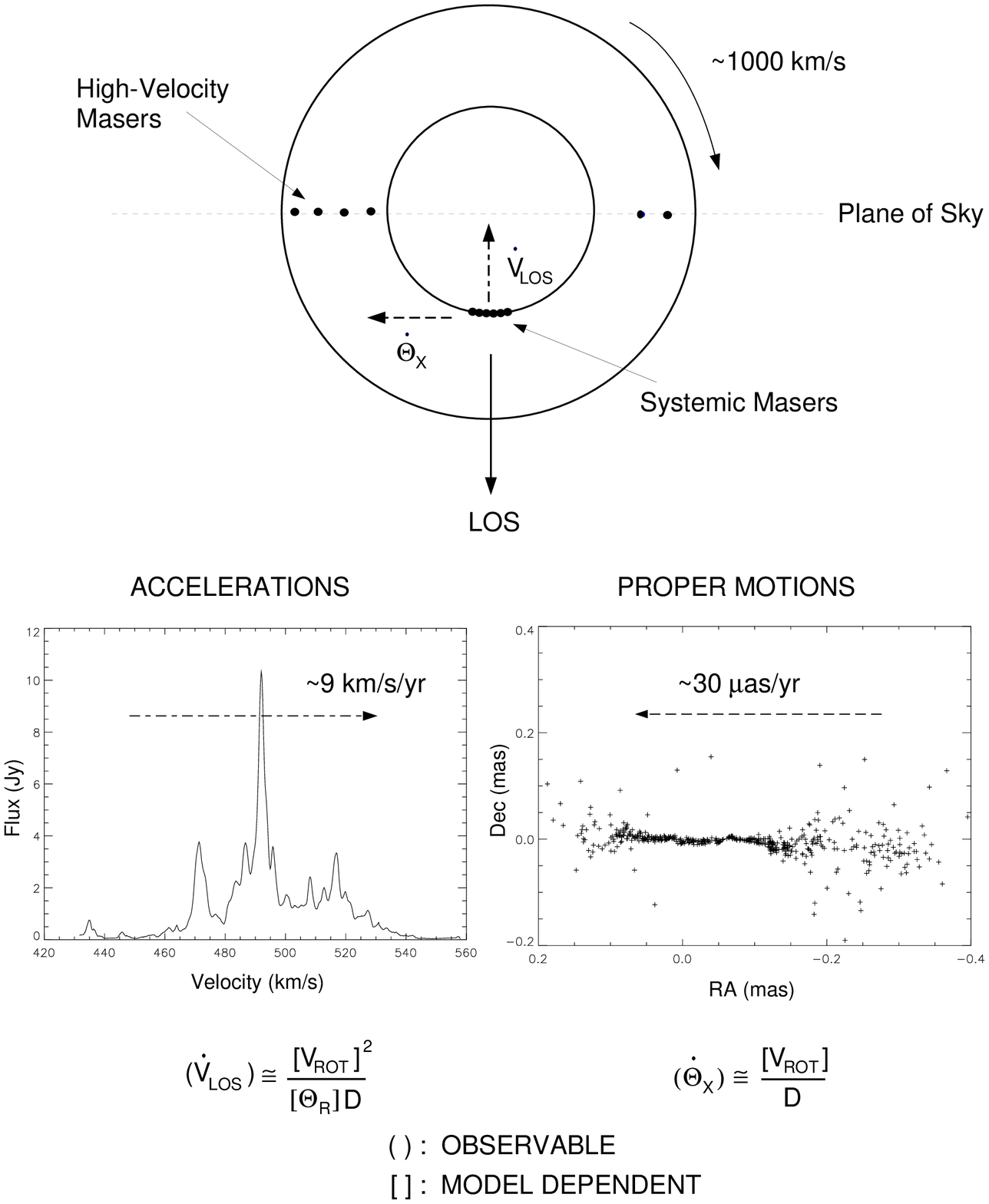}
\caption {The top panel is a schematic representation of the sub-parsec molecular
disk in NGC 4258.  The bottom panels show the expected acceleration and proper
motion vectors superposed on actual VLBA data of the systemic masers taken in May of
1995.}
\end{figure}

\section{Masers and Cosmology?}

Thus, the nuclear maser in NGC 4258 has yielded a purely 
geometrical distance with fractional uncertainty of 4\% that is completely 
independent of all the usual rungs of the extragalactic distance ladder.
While this result will undoubtedly play an important role in calibrating other
primary and secondary distance indicators, it will be of limited utility
in directly constraining any cosmological parameters due to the relative 
proximity of NGC 4258. The potential scientific
payoffs from discovering an NGC 4258-like object at large distance, in the 
Hubble flow, are considerable: a 5\% distance to such a maser could place 
significant constraints on both the Hubble constant and the deceleration 
parameter.  In the remainder of this contribution, I describe why such
high-redshift masers may, at least in principle, be detectable, and I 
discuss a program we have implemented to uncover them in practice. 

The top portion of Figure~2 shows the inner 10 pc of a typical high-luminosity 
($L\sim10^{45-47}$ erg s$^{-1}$) AGN, as depicted in standard models. Between 
0.1 and 1 pc
lie the broad and intermediate line regions (BLR and ILR), characterized by 
forbidden optical line widths of $\sim7000$ and $\sim2000$ km s$^{-1}$ and 
number densities of $10^{12.5}$ and $10^{10}$ cm$^{-3}$, respectively (Baldwin 
1997).  This gas is both too hot and too dense to support water masers, which 
require number densities between $10^{8}$ and $10^{10}$ cm$^{-3}$ and 
temperatures between 200 and 1000 K. The outer edge of the ILR is probably 
caused by the formation of dust beyond 1 pc (Netzer \& Laor 1993). 
This dust shields the gas from the central ionizing radiation and 
lowers the optical emissivity of the gas, resulting in the `dead zone', which 
is largely devoid of optical emission lines, and hence remains essentially
unstudied.  However, the shielding of the dust tends to {\it promote} the 
creation of water outside of 1 pc. Furthermore, the presence of dust may help 
maintain an inverted population of water by absorbing the IR line photons 
that would otherwise tend to thermalize the population (Collison \& Watson 1995).
Finally, Rees, Netzer, \& Ferland (1989) estimate a number density of
$10^{8-9}$ cm$^{-3}$ at the outer edge of this zone (10 pc).  {\it Thus, the 
standard model for luminous AGN suggests that conditions suitable for water
masers should exist between 1 and 10 pc from the central engine.}

The flux density at which an amplifying maser saturates ($F_{s}$) is given by
(c.f. Haschick {\it et al.} 1990)
\begin{equation}
F_{s}\simeq 12(D_{m}/0.15\mbox{ pc})^{2}(D/7.3\mbox{ Mpc})^{-2} \mbox{~~~Jy}, 
\end{equation} 
where $D_{m}$ is the 
distance between the maser and the background source, and $D$ is the distance 
to the maser. Here, we postulate that this saturation flux density is a 
representative flux 
for an amplifying maser, provided the gain requirements are not too severe (this 
is indeed the case in NGC 4258, where both $D_{m}$ and $D$ are known to high 
precision; Herrnstein {\it et al.} 1997).  Above, we have argued that water
masers in high-luminosity AGN are likely to be located 1 -- 10 pc from the 
central engine and/or the radio core, the most promising 
reservoirs of background micorowave seed photons. 
{\it The saturation flux for a 
maser at $D_{m}=10$ pc is $2.8(D/\mbox{Gpc})^{-2}$ Jy. Such a maser should be 
easily detectable well into the Hubble flow.} 

\begin{figure}[p]     
\plotone{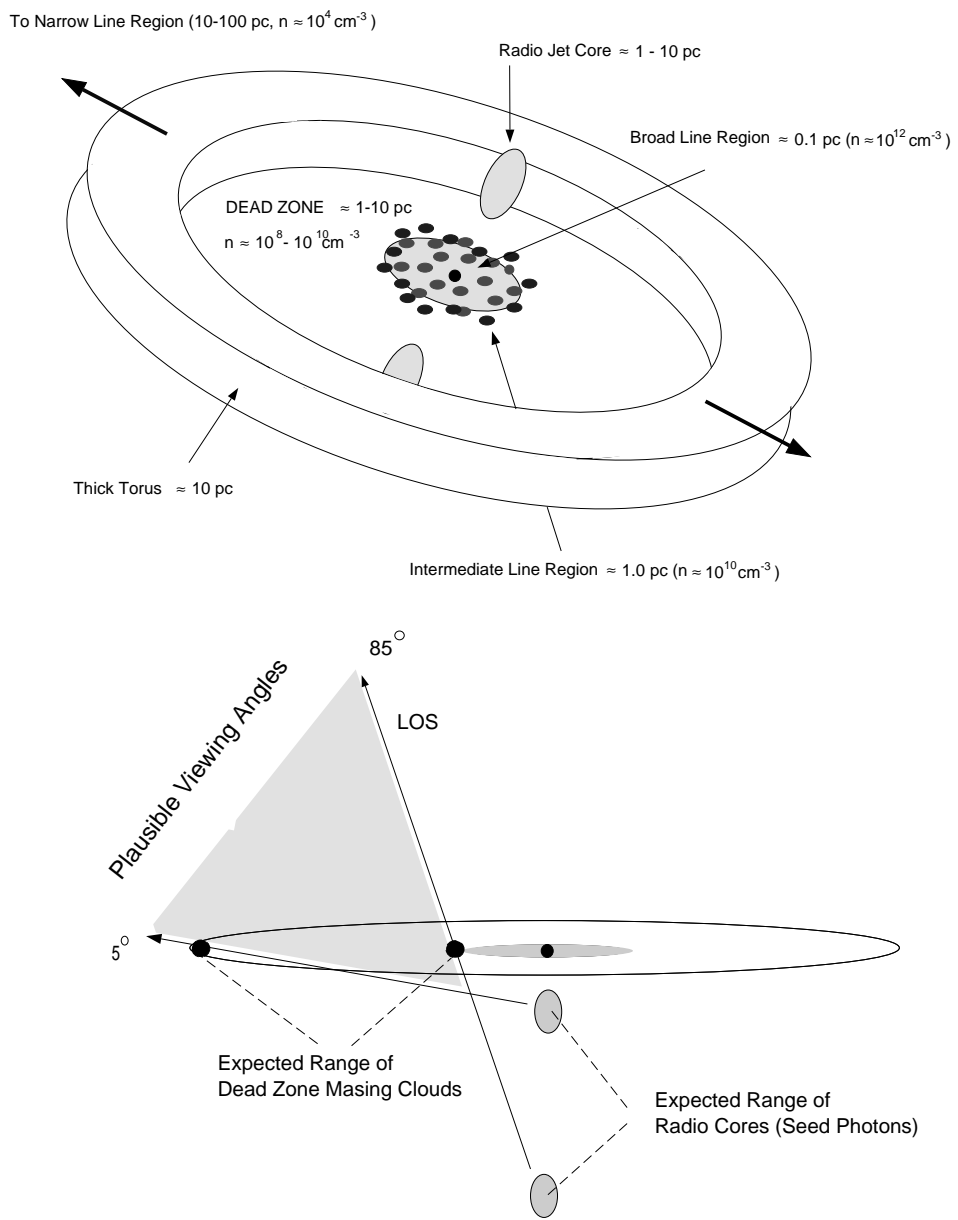}
\caption {Schematic representation of a high-luminosity active galactic nuclei.  See
text for details.}
\end{figure}

The proposed high-redshift masers achieve their enormous apparent luminosities 
through miniscule beaming angles, and an obvious concern is that they will be 
extremely difficult to find.  However, if the maser-cloud filling factor in the 
'dead zone' of Figure~2 is reasonably large, then such concerns may not be 
justified. The probability of realizing the necessary alignment between any 
one of the many foreground maser clouds and a background continuum source is 
much greater than is implied by the extremely narrow beam angle of any particular 
maser. Each maser cloud beams in a slightly different direction, and only one 
of these beams must lie along our LOS. We note that standard models for 
core-jet radio emission predict that the 22 GHz core can lie anywhere from 1 
to 10 pc from the nucleus (Blandford \& K$\ddot{\mbox{o}}$nigl 1979). This 
relatively large spread, combined with the considerable extent of the `dead 
zone', indicates that detectable high-redshift masers may not be limited to 
AGN of any particular orientation.

The apparent luminosities of the proposed high-z masers are vastly larger than 
those of all known nearby megamasers, and it is natural to ask why we do not see 
any {\it nearby} apparently ultra-luminous masers.  Essentially all of the $\sim500$ 
galaxies searched for masers to date have been nearby, relatively low-powered AGN 
such as LINERS or Seyfert 2s (Braatz 1996).  As a result, the known, nearby masers 
invariably reside in relatively low-powered systems for which $D_{m}$, and 
therefore the saturation flux, is expected to be much smaller than in the 
high-powered, high-redshift systems. For example NGC 4258, with a bolometric 
luminosity of about $10^{42}$\,erg\,s$^{-1}$ (Herrnstein {\it et al.} 1998a), is 
known to have a $D_{m}$ of about 0.15 pc (Herrnstein {\it et al.} 1997).  In 
the context of the archetypal high-luminosity AGN of Figure~2, this result
is consistent with the crude generalization that AGN sizes ought to scale as 
$\sim L^{1/2}$. 

\section{A High-Redshift Water Maser Search}

This spring, we will use the VLBA to search for water masers in a sample of 114 
moderate- and high-redshift AGN.  This sample, generated using the NASA 
Extragalactic database (NED), is comprised mostly of QSOs and radio-loud AGN, and 
consists of all those sources in either the USNO geodesic survey or the VLBA 
calibrator survey above $-30^{\circ}$ dec with (1) redshifted 22 GHz water 
maser emission falling within the VLBA U-, X-, or C-bands, and (2) 5 GHz 
fluxes greater than 20 mJy.  For the reasons described above, we will observe 
all sources that satisfy these criteria, irrespective of the likely orientation of 
the AGN.  The breakdown of sources by redshift and band is shown in Table~1.  We 
will spend approximately 10 minutes on each source, and cover the entire sky in a 
continuous 24-hour track, taking advantage of the excellent frequency agility of
the VLBA.

Table~1 lists the velocity coverage and resolution in the rest frame of the host 
galaxy for each band. The large redshifts and 64-MHz bandwidth conspire to give 
extremely broad velocity coverage.  The VLBA correlator also enables us to maintain 
adequate velocity resolution even for sources at $z=3.5$.  The combination of 
$>1000$ km s$^{-1}$ bandwidths and $<1$ km s$^{-1}$ resolutions is unique amongst 
maser searches, and is largely due to the VLBA correlator.  We will self-calibrate 
on the continuum emission, and generate continuum snapshots for the entire sample.  
This collection of snapshots, unique in that they are at the same frequency in the 
{\it rest frame} of the objects, will be made available to the public.  The phases 
from the continuum self-calibration will be applied to the spectral data to stabilize 
the interferometer.  Finally, we will generate image cubes (limited to several beams 
per spatial dimension) and search for maser emission in the synthesized spectra.  
The baseline-based thermal noises in the fifth column of the table confirm the 
feasibility of the continuum self-calibration. The sixth column gives the expected
image thermal noises in ten minutes, and averaged across a 1 km s$^{-1}$ maser 
feature.  {\it The anticipated 10-15 mJy thermal noises are as good or better than 
any previous water maser search performed to date.  Furthermore, because the VLBA
is a cross power instrument, we minimize the deleterious effects of standing 
waves on the detection of broadened maser emission.}

Finally, we note that Barvainis and collaborators are also searching for 
high-redshift water masers using the Effelsberg 100-meter telescope.  Preliminary
results from this parallel effort are presented elsewhere in these proceedings.

\begin{table}[hp]
\begin{center}
\caption{Velocity resolutions and sensitivities} \vspace{0.4cm}
\label{tb:tab}
\begin{tabular}{cccccccc} \hline\hline
 Band  &      f       &     $z$    & $N^{(1)}$  & $\Delta v^{(2)}$ & $v_{s}^{(3)}$ & $\Delta S_{2m}^{(4)}$  & $\Delta S_{l}^{(5)}$  \\
       &    (GHz)     &           &    &  (km/s)    &  (km/s) &    (mJy)         &     (mJy)       \\ \hline
   U   & 12.0 - 15.4  & 0.44-0.85 & 82 &  1400      &   0.35  &      6           &     15          \\
   X   & 8.0 - 8.8    & 1.52-1.78 & 25 &  2300      &   0.56  &      4           &     10          \\
   C   & 4.6 - 5.1    & 3.35-3.83 &  7 &  4000      &   1.0   &      4           &     15          \\
\hline
\end{tabular}
\end{center}
\tablenotetext{1}{Number of sources within specified range of $z$ with: (1) $F_{5 GHz} > 20$ mJy, 
(2) $\delta > -30^{\circ}$, and (3) VLBI position.}
\tablenotetext{2}{Velocity coverage {\it in the host galaxy} for 64 MHz total bandwidth.}
\tablenotetext{3}{Velocity resolution {\it in the host galaxy} for 512 channel spectra.}
\tablenotetext{4}{Continuum sensitivity along a single VLBA baseline in 2 minutes (64 MHz, 1 bit sampling).}
\tablenotetext{5}{Image sensitivity in maser line, for a 10-minute integration and assuming 1 km s$^{-1}$ linewidth.}

\end{table}

\acknowledgements

The National Radio Astronomy Observatory is operated by Associated 
Universities, Inc, under cooperative agreement with the National Science Foundation.
The NASA/IPAC Extragalactic Database (NED) is operated by the Jet Propulsion
Laboratory, California Institute of Technology, under contract with the National
Aeronautics and Space Administration.
I thank A. Beasley, L. Greenhill, A. Loeb, and J. Moran for helpful discussions.

\end{document}